\documentclass[11pt]{article}
\usepackage{moriond,epsfig}

\def\be{\begin{equation}}
\def\ee{\end{equation}}
\def\bea{\begin{eqnarray}}
\def\eea{\end{eqnarray}}

\def\la{Lyman-$\alpha$\,}
\begin{document}
\vspace*{4cm}
\title{THE LYMAN-$\alpha$ FOREST ACCORDING TO LUQAS}

\author{M. VIEL$^{1}$, M. HAEHNELT$^{1}$, T.-S. KIM$^{1}$,
B. CARSWELL$^{1}$, S. CRISTIANI $^2$, A. HEAVENS $^3$, L. HERNQUIST $^{4}$. S. MATARRESE
$^5$, V. SPRINGEL$^6$}

\address{$^1$ Institute of Astronomy, Madingley Road, CB3 0HA Cambridge, UK
\\
$^2$ INAF-Osservatorio Astronomico di Trieste, via G.B. Tiepolo 11,
I-34131 Trieste, Italy\\
$^3$ Institute for Astronomy, University of Edinburgh, Blackford Hill,
Edinburgh EH9 3HJ\\
$^4$ Harvard-Smithsonian Center for Astrohpysics, 60 Garden Street,
Cambridge, MA 02198, USA\\
$^5$ Dipartimento di Fisica `Galileo Galilei' e INFN Sez. di Padova, via Marzolo 8, I-35131 Padova, 
Italy \\
$^6$ Max-Planck-Institut f\"ur Astrophysik, Karl
Schwarzschild-str. 1,Garching bei M\"unchen, Germany
\\
}

\maketitle\abstracts{We  use the LUQAS sample of 27 high resolution, high 
signal-to-noise QSO absorption spectra (Kim et al. 2004) and the results from 
Croft et al. (2002) together with a suite of high-resolution
hydro-dynamical   simulations run with the GADGET-II code, to infer the
linear dark matter power spectrum on scales of $\sim 0.3-30$ Mpc at
$z=2.125$ and at $z=2.72$.}

\section{Introduction}

The Ly$\alpha$ forest in QSO absorption spectra has been used to probe dark matter power
spectrum  by a number of authors (e.g. Croft et al. \cite{croft}; Gnedin \&
Hamilton \cite{gnedin}; Viel, Haehnelt \& Springel \cite{viel4} and
references therein).  The  Ly$\alpha$ forest is sensitive to
fluctuations in the DM density on scales of a few Mpc. It can thus be
used to investigate the  possible cut-off of the DM spectrum which is 
expected if the dark matter were warm dark matter, it can give
constraints on the  matter fraction in neutrinos and allows to
investigate the gravitational growth of structure and
possibly the redshift evolution of dark energy (Viel et
al. \cite{viel1}; Mandelbaum et al. \cite{mandelbaum}; Lidz et
al. \cite{lidz}). Verde et al.  \cite{verde} combined the Ly$\alpha$
forest results of Croft et al. with the CMB results of WMAP  and
concluded that there is 
evidence for a running spectral
index and for a tilt in the primordial power spectrum. However, a
recent analysis by Seljak et al.  \cite{seljak} argued that 
the errors are larger and the effective optical depth is smaller 
than that assumed by Verde et al.   Seljak et al. concluded  that there
is no evidence for a running spectral index or for a tilt of  the
primordial power spectrum.  
We briefly discuss  here the flux power spectrum of the 
LUQAS sample (Kim et al. \cite{kim}), a set of high resolution  QSO
absorption spectra, and compare it to  the recently published flux
power spectrum obtained for a large sample of low-resolution 
SDSS spectra (McDonald et al. \cite{mcdonald}). We will further discuss  
the linear dark matter power spectrum which Viel et al. \cite{viel4} obtained from
the LUQAS flux power spectrum   using the effective bias method proposed by Croft et
al. \cite{croft}. Implications for the rms fluctuation amplitude
$\sigma_{8}$ and the primordial index $n$ are also discussed.

\begin{figure}
%\rule{5cm}{0.2mm}\hfill\rule{5cm}{0.2mm}
%\vskip 2.5cm
%\rule{5cm}{0.2mm}\hfill\rule{5cm}{0.2mm}
\center\psfig{figure=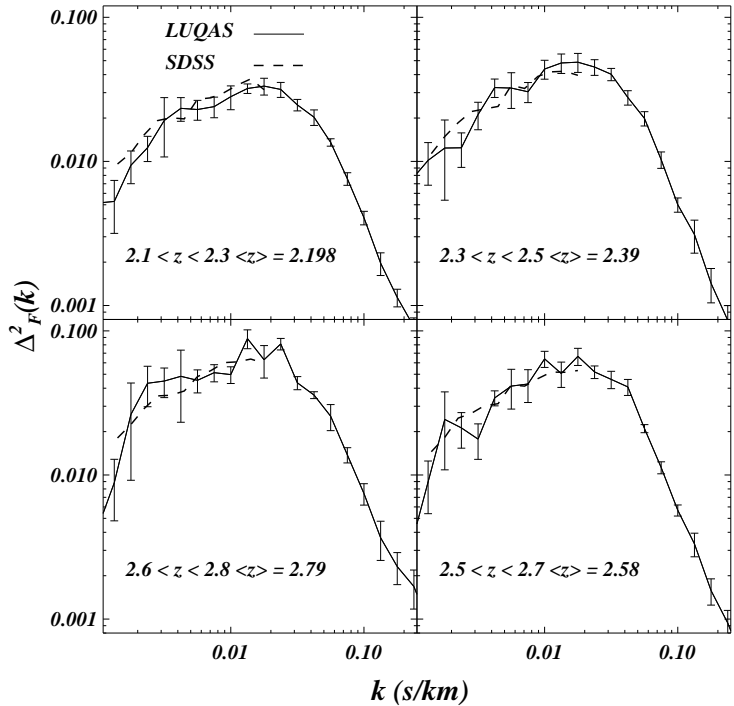,height=10.42cm,width=10.42cm}
\caption{Comparison between the flux power spectrum of the LUQAS
sample (continuous line with error bars) and the recent results for a large sample of SDSS
spectra by  McDonald et al. 2004 (dashed line). Note that the SDSS
data refer to their table 3 including the background contribution.}
\end{figure}

\section{The LUQAS sample}
The LUQAS sample \footnote{Tables of the flux power spectrum are available at: {\rm
www.ast.cam.ac.uk/$\sim$rtnigm/luqas.html}. Please note that the
k-values of tables 3 and 5 of Kim et al. \cite{kim} had been erroneously shifted
by half a bin-size in log k (erratum in
press). Corrected tables can also be found in astro-ph/0308103.}
(Large Sample of UVES Quasar Absorption Spectra) consists of spectra
of 27 QSOs taken with the Ultra-Violet Echelle Spectrograph (UVES) on
VLT. Most of the spectra have been taken as  part of the Large  ESO Observing programme  UVESLP
(P.I.: J. Bergeron). The median redshift of the sample is $z=2.25$ and
the total redshift  path is$\Delta z=13.75$. The typical
signal-to-noise ratio is $\sim 50$ and the pixel size is 0.05 \AA. For
a more detailed description of the sample and the data reduction  we refer to Kim et al.
\cite{kim}.  In Figure 1 we compare the 1D flux power spectrum of four 
subsamples of LUQAS at different redshifts (continuous curves) with the
flux power spectrum obtained by McDonald et al. \cite{mcdonald}  
from a large sample of low-resolution SDSS spectra (dashed
curves).

There is reasonable good agreement between the two estimates of the
flux power spectrum over a wide range of wave-numbers, in all  four
redshift bins. 
Note that due to the significantly higher resolution the LUQAS sample can 
constrain the matter power spectrum all the way to the thermal cutoff 
($k>0.01$ s/km). At these small scales the estimate of the  SDSS
sample is affected by its low resolution.  The statistical errors of the SDSS sample are 
significantly smaller but the systematic uncertainties are
larger than the statistical errors (Kim et al. \cite{kim}, McDonald et
al. \cite{mcdonald}). The  flux bispectrum of the LUQAS sample has been presented 
in Viel et al. \cite{viel3}. Note that the large discrepancy between
the two samples claimed in McDonald et al. was mainly due to an 
an erroneous shift of the  flux power spectrum by half a bin-size in
log k in Kim et al. (2004, erratum in press).

\section{Inferring the linear dark matter power spectrum}

The method we use to infer the linear dark matter power spectrum has
been proposed by Croft et al. \cite{croft} (but see Gnedin \& Hamilton
\cite{gnedin}). 
It uses numerical simulations to calibrate the
relation between flux power spectrum and matter power spectrum.  It
then assumes that the flux power spectrum $P_F(k)$ at a given
wavenumber $k$ depends linearly on the linear real space matter power
spectrum $P(k)$ at the same wavenumber and that both can be related by
a simple bias function $b(k)$ as $P_F(k)=b^2(k) \,P(k)$.  

We have run a suite of hydro-dynamical simulations with varying cosmological
parameters, particle numbers, resolution, boxsize and thermal
histories using a new version of the parallel TreeSPH code {\small
GADGET}, in order to explore the different systematics and statistical
errors.

\begin{figure}
\psfig{figure=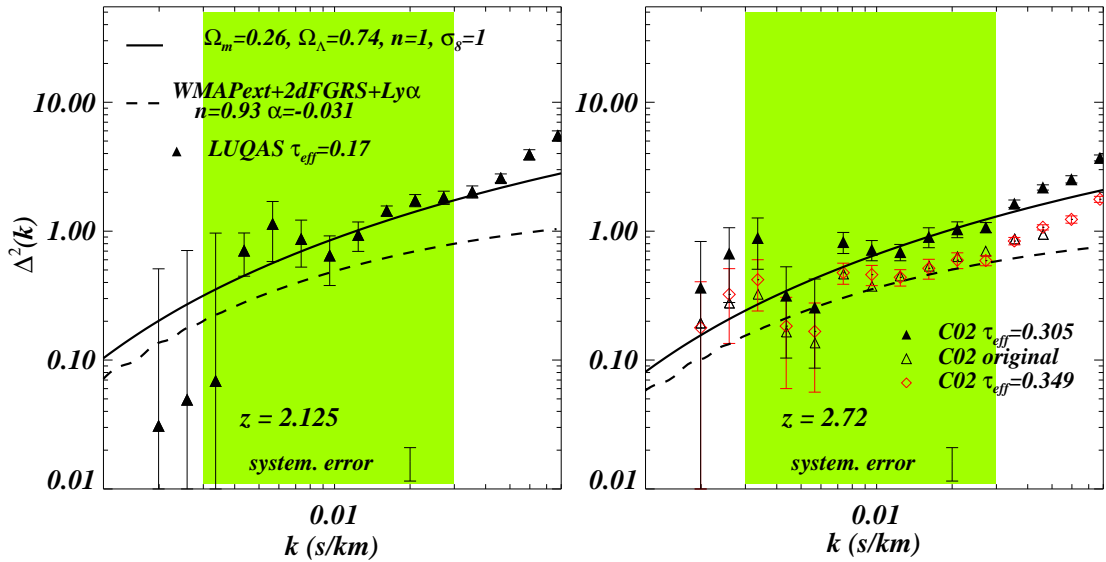,height=8.0cm,width=16cm}
\caption{The recovered linear dark matter power spectrum at two
different redshifts. At $z=2.125$ using the LUQAS sample (left panel)
and at $z=2.72$ using the Croft et al. (2002) sample (right panel). The
filled triangles have been obtained with an effective optical depth
derived from high resolution spectra and from hydro-simulations with $T_0=10^{4.15}$ K and $\gamma=1.6$. In the right panel we overplot
the original result from Croft et al. (empty triangles) and a
re-analysis of their data with the effective optical depth they assumed
(diamonds). Continuous black line shows a model which is a good fit to our
data. The dashed line represents
a model with a running spectral index with best fitting values
suggested by the WMAP experiment (Table 10, Spergel et al. 2003). 
The yellow rectangles indicate the range of wavenumbers used in our analysis.}
\end{figure}

In Figure 2 we show the results of our analysis in terms of
$\Delta^2(k)= P(k)\,k^3/(2\pi^2)$ for the LUQAS sample (left panel) at
$z=2.125$ and for the Croft et al. sample (right panel) at
$z=2.72$. The filled triangles represent the recovered linear dark
matter power spectrum using values of $\tau_{\rm eff}$ suggested by
high-resolution spectra.

The  dashed curves in Figure 2 represent  the best fitting running
spectral index model of Spergel et al. \cite{spergel} which is a fit
to WMAP, CBI, ACBAR, 2dFGRS
and the \la forest data, while the continuous line is a good fit to our
data points.

In order to minimise uncertainties due to continuum fitting and metal lines,
and to avoid dealing with the problematic thermal cut-off at small
scales, we have  only used the range of wavenumbers $0.003 < k  \,{\rm
(s/km)} <0.03$ for our quantitative analysis, which corresponds roughly 
to scales 0.3-30 Mpc.

The main results can be summarized as follows.  With the same assumptions for
effective optical depth, density-temperature relation, and cosmology,
our inferred linear matter power spectrum (empty diamonds right panel)
agrees very well with that inferred by Croft et al. \cite{croft} (empty
triangles right panel).

We confirm previous results that the inferred rms amplitude of density
fluctuations depends strongly on the assumed $\tau_{\rm eff}$. It
increases by 20\% if we assume an optical depth of $\tau_{\rm eff}=
0.305$ a value indicated form high-resolution absorption spectra. We
find, however, a dependence on $\tau_{\rm eff}$ which is weaker than
that of Croft et al. \cite{croft} and stronger than that of Gnedin \&
Hamilton \cite{gnedin} . The decrease of the amplitude of the flux
power spectrum between $z=2.75$ and $z=2.125$ is consistent with that
expected due to the decrease of $\tau_{\rm eff}$ and the increase of
the amplitude of matter power spectrum due to gravitational growth.

Our estimate of the systematic uncertainty (error bars in the bottom
part of the two panels in Figure 2) of the rms fluctuation amplitude of
the density ($\sim 14.5 \%$) is a factor 3.5 larger than our estimate
of the statistical error ($\sim 4 \%$).  The systematic uncertainty is
dominated by the uncertainty in the mean effective optical depth and by
the uncertainties between the numerical simulations of different
authors. Reducing the overall errors will thus mainly rely on a better
understanding of a range of systematic uncertainties.

By combining the  constraint on the amplitude of the DM power spectrum 
on large scale from COBE (assuming that there is no contribution
from tensor fluctuations) with the constraint from  the
high-resolution \la forest data on small scales we obtain $n= 1.01
\; (\Omega_{\rm 0m}h^2/0.135)^{-0.35}\; \pm 0.02 {(\rm statistical)}\pm
0.06{( \rm systematic)}$ for the spectral index. The corresponding rms
fluctuation amplitude is, $\sigma_8 = 0.93 \pm 0.03 {(\rm
statistical)}\pm 0.09{(\rm systematic)}$.  

Thus, for values of the mean optical depth favoured by high-resolution
spectra, the inferred linear power spectrum is consistent with a
$\Lambda$CDM model with a scale-free ($n=1$) primordial power
spectrum.

\section*{Acknowledgements}
This work is supported by the European Community
Research and Training Network ``The Physics of the Intergalactic
Medium''. MV acknowledges support from a EU grant for young scientists
and from PPARC.  The simulations
were run on the COSMOS (SGI Altix 3700) supercomputer at the
Department of Applied Mathematics and Theoretical Physics in Cambridge
and on the Sun Linux cluster at the Institute of Astronomy in
Cambridge. COSMOS is a UK-CCC facility which is supported by HEFCE and
PPARC.

\end{document}